# Detection of deep-subwavelength dielectric layers at terahertz frequencies using semiconductor plasmonic resonators

Audrey Berrier,[1,*] Pablo Albella,[3] M. Ameen Poyli,[3] Ronald Ulbricht,[2] Mischa Bonn,[2] Javier Aizpurua,[3] Jaime Gómez Rivas[1,4]

[1]*FOM Institute AMOLF, Centre for Nanophotonics, c/o Philips Research Laboratories, HTC4, 5656AE  Eindhoven, The Netherlands*
[2]*FOM Institute AMOLF, Science Park 104, 1098XG Amsterdam, The Netherlands*
[3]*Center for Materials Physics CSIC-UPV/EHU and Donostia International Physics Center DIPC, Paseo Manuel Lardizabal, 20.018 Donostia- San Sebastian, Spain.*
[4]*COBRA Research Institute, Eindhoven University of Technology, P.O. Box 513, 5600 MB Eindhoven, The Netherlands .*

[*]*a.berrier@amolf.nl*

**Abstract.** Plasmonic bowtie antennas made of doped silicon can operate as plasmonic resonators at terahertz (THz) frequencies and provide large field enhancement close to their gap. We demonstrate both experimentally and theoretically that the field confinement close to the surface of the antenna enables the detection of ultrathin (100 nm) inorganic films, about 3750 times thinner than the free space wavelength. Based on model calculations, we conclude that the detection sensitivity and its variation with the thickness of the deposited layer are related to both the decay of the local THz field profile around the antenna and the local field enhancement in the gap of the bowtie antenna. This large field enhancement has the potential to improve the detection limits of plasmon-based biological and chemical sensors.

## 1. Introduction

Electromagnetic radiation in the terahertz (THz) regime has a widely recognized potential for sensing owing to its capability to couple to various low-energy resonances of matter, including rotational and vibrational motion of molecules, as well as charge carriers and quasi-particles, such as plasmons, in semiconductors [1,2]. The ability to interrogate specific fingerprints of particular materials renders THz-based approaches promising for the detection and recognition of strategic substances such as metals, explosives, gases, organic or biological substances [3,4]. However, the wavelength of the THz radiation, *e.g.*, 375 μm at 0.8 THz, makes the access to nanometric sensing volumes challenging. Conventional THz spectroscopy makes use of large amounts of matter (requiring flow cells of the order of 1 m for gas spectroscopy [5,6] and pellets of some mm for solids [3]). The quest for finding mechanisms that enhance the signal of terahertz radiation in small volumes, hence reducing the amount of matter needed for THz spectroscopy, is therefore a natural drive in this field.

Metal plasmonic structures at optical and near infrared frequencies have proven successful at enhancing the signal of several spectroscopic techniques including surface-enhanced Raman spectroscopy (SERS) [7-9] and surface-enhanced Infrared absorption (SEIRA) [10-12]. The plasmon-induced enhancement of the sensitivity is related to the resonant behavior of the plasmonic structures that host the target substances and the corresponding field enhancement produced in their proximity at those energies. For example, monolayers of organic substances were successfully sensed at infrared frequencies using this approach [12,13].

At THz frequencies however, the potential of resonant structures to improve sensing sensitivity has remained relatively unexplored. Conventionally, thick layers of substance are used for material characterization [14], where waveguiding can be used to increase the THz radiation-matter interaction [15]. The use of THz plasmonic structures for the detection of thin layers has recently been demonstrated [16-19]. Organic films of 500 nm thick were detected using THz metamaterial structures [18]. The detection of a polystyrene layer of thickness $\lambda/1000$ has been reported [17]. THz sensors with improved sensitivity capabilities despite very small sample volumes have potential applications in areas as diverse as environment monitoring, lab-on-chip for point-of-care monitoring and medical diagnosis.

Here, we show that bowtie antennas made of doped silicon operating as plasmonic resonators at THz frequencies are a versatile platform for thin film detection. Compared to metallic resonators, semiconductor-based structures are easily tunable [20-22] and operate in a regime where the skin depth of the material is larger, *i.e.*, the impedance is lower, and hence the coupling to surface plasmons is more pronounced. A structure such as a bowtie made of doped silicon provides large field confinement and enhancement in the region of its gap at THz frequencies [21]. When an inorganic thin film is deposited on top of the bowtie antenna, the area around the gap of the antenna thus provides an enhanced THz field that results in an enhanced interaction of the terahertz radiation with the deposited ultrathin inorganic films, allowing for THz spectroscopy in very small volumes. We experimentally demonstrate the *in-situ* detection of films that are orders of magnitude thinner than the wavelength using doped silicon bowtie antennas. In particular, we show that semiconductor bowtie antennas operating at THz frequencies allow the sensing of thin inorganic films with a layer thickness as small as $\lambda/3750$. THz plasmonic antennas thus provide a platform for THz spectroscopy in much smaller volumes, for which the tested material quantity is reduced significantly.

## 2. Experimental

The structures presented here have been fabricated using conventional micro-fabrication techniques. Silicon-on-insulator (SOI) wafers were implanted with arsenic atoms followed by an activation step at 1050°C, resulting in a carrier concentration of about $(6\pm3)\times10^{19}$ cm$^{-3}$. It has been shown [20,21] that, owing to their increasing value of permittivity with carrier concentration, doped silicon is a suitable material for plasmonics at THz frequency. The SOI wafer was subsequently bonded to a quartz wafer by BCB (benzocyclobutene) bonding resulting in a BCB layer of a few microns between the Si and quartz substrates. The back silicon substrate and the silica buffer layer were removed by wet chemical etching using KOH and HF solutions, respectively. The final vertical structure consists of a quartz wafer acting as substrate, a BCB layer, and a 1.5 μm thick doped silicon layer. The bowtie structures were defined by conventional optical lithography, followed by reactive ion etching of the silicon layer using the photoresist as a mask and subsequent photoresist stripping. Similarly to Ref. 21, the bowtie antennas are arranged in a non-periodic pattern in order to avoid any collective behavior due to periodicity. The SiO$_2$ and TiO$_2$ layers were subsequently deposited on the substrate that holds the patterned bowtie structures using plasma enhanced chemical vapor deposition (PECVD), which allows a conformal deposition around the antennas. Figure 1(a) shows a schematic drawing of the silicon bowtie obtained with use of this technique together with the deposited layer. Samples with oxide layer thicknesses between 100 nm and 2 μm were obtained following this procedure.

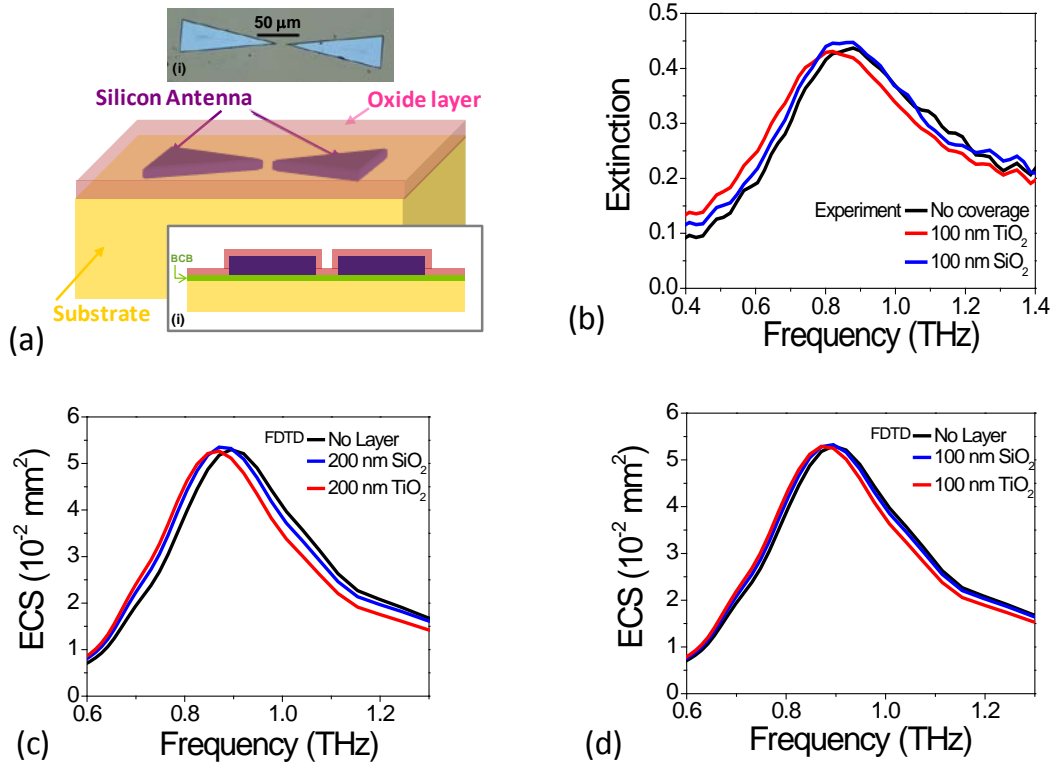

Fig. 1. a) Schematic drawing of a silicon bowtie antenna covered with a TiO$_2$ conformal layer. The inset (i) shows an optical microscopy image of a fabricated silicon bowtie antenna, the inset (i) represents a schematic drawing of the vertical cross section in the middle of the bowtie antenna; b) Far-field extinction of a collection of bowtie antennas without coverage (black line), with a 100 nm TiO$_2$ layer (red line) and with a 100 nm SiO$_2$ layer (blue line); c) FDTD calculated extinction cross section (ECS) for an individual doped silicon antenna with a 200 nm layer of SiO$_2$ (blue line) and TiO$_2$ (red line) deposited on the antenna surface; d) FDTD calculated extinction cross section (ECS) for an individual doped silicon antenna with a 100 nm layer of SiO$_2$ or TiO$_2$.

The terahertz response of the samples is experimentally characterized using a terahertz time domain spectrometer (THz-TDS) at room temperature. The set-up was purged with N$_2$ gas to remove water vapor, which strongly absorbs THz radiation. The bowtie antennas were designed to resonate around 0.8 THz [21]. Ref. 21 demonstrates the ultrafast activation of the plasmonic resonance of undoped silicon bowtie antennas by photogeneration of free carriers using an optical pulse. In this work, we use doped silicon to allow localized surface plasmons to be excited without the need of optical pumping. Experimentally, we define the extinction of the sample as 1-$T$, where $T$ is the transmission through the sample normalized to the transmission through the bare substrate. Defined in this way, the extinction captures both effects of absorption and scattering. The extinction of the bowtie structure in the presence and absence of the oxide layer is shown in Figure 1(b). The origin of the peak in extinction is attributed to the excitation of localized surface plasmon polaritons (LSPP) in the silicon bowtie structure. LSPPs are resonances due to the collective excitation of the free charge carriers in the material and depend both on material properties (*e.g.*, charge carrier density and carrier mobility) and on the geometry of the structure [23]. The measured extinction spectrum of the structures covered with a 100 nm $SiO_2$ layer (red line) is shifted with respect to the non-covered structure (black line) indicating the ability of this method to detect 100 nm thick layers of a material with refractive index $n\sim1.95$ at 1 THz [24]. This corresponds to a thin film of thickness $\lambda/3750$. Furthermore, the magnitude of the resonance shift depends on the refractive index of the layer. We demonstrate this by depositing a 100 nm layer of a different material ($TiO_2$) on a similar array of bowtie antennas ($n\sim9$ at 1 THz for $TiO_2$ [24]). It is evident from Fig. 1(b) that the shift of the bowtie resonance is larger for $TiO_2$ than for $SiO_2$, as expected from their respective values of the refractive indexes.

3. **Comparison with simulations**

Finite-difference time-domain (FDTD) simulations of the extinction cross section were performed for the system used in the experiment, using a commercial package to solve Maxwell's Equations (Lumerical). The simulations are performed for an isolated bowtie structure surrounded by perfectly matched layers, and the discretisation of space used a mesh size of 200 nm in the plane of the antenna and a mesh size of 10 nm in the direction normal to the antenna, which was tested to provide convergence of the results without loss of any significant physical detail. The material properties for silicon are described using the Drude model with the mobility dependence given in Ref 25. The carrier concentration used for the simulations is $8\times10^{19} cm^{-3}$, to be consistent with the experimental results. In particular, the values of the permittivity at 0.8 THz are for doped silicon $\varepsilon\approx-7.5\times10^2+i\cdot3.1\times10^2$, and for gold $\varepsilon\approx-1\times10^5+i\cdot1\times10^6$. The response of the bowtie antennas to the inorganic layer coverage can be reproduced by the FDTD simulations, as shown in Fig. 1(c) and (d) where the extinction cross section (ECS) of a single antenna is shown for different layer thicknesses. The experimental results shown in Fig. 1(b) represent the total extinction experimentally extracted from the transmission through a collection of bowtie antennas. The magnitude of the shift of the resonance depends both on the refractive index of the layer and on its thickness. As expected, either larger refractive indices or increased layer thickness result in larger spectral shifts of the resonance. The values for the theoretical spectral shifts shown in Fig. 1(c) and (d) demonstrate the capacity of this structure to resolve dielectric layers as thin as 100nm for both $SiO_2$ and $TiO_2$. The shift induced by a 100 nm layer of $TiO_2$ is larger than that of 100nm $SiO_2$ due to the higher refractive index of $TiO_2$. As discussed in the following, the resonance shift depends both on layer thickness and on refractive index. This indicates that a layer of 30nm of $TiO_2$ would induce a shift similar to that of 100nm $SiO_2$. Within the signal-to-noise ratio of our measurements, this suggests that films as thin as 30 nm of $TiO_2$ should be readily detectable. In terms of layer thickness relative to the wavelength is a 10-fold improvement relative to previous approaches using THz plasmonics [17].

We analyze now the spectral dependence on the layer thickness. Figure 2(a) displays the experimental extinction spectra of the doped silicon bowtie structure covered by a $SiO_2$ layer of varying thickness, ranging from 100 nm to 2 μm. As the thickness of the $SiO_2$ layer increases, the resonance peak red-shifts and the magnitude of the extinction peak increases in accordance with the increase of the volume of the antenna. In the limit of an infinitely thick layer the peak shift is roughly of 0.4 THz. This red-shift can be qualitatively understood by dielectric screening of the elements of the bowtie antenna [23] and is reproduced by FDTD simulations, as observed in Fig. 2(b). We note that due to under-etching during the fabrication process, the distance separating the two monomers of the bowtie antennas is relatively large (around 20 μm). The coupling between the two triangular structures that constitute the bowtie for this relatively large separation is weak. As a result, the response of the bowtie is similar to that of isolated triangular structures, *i.e.*, an uncoupled system. Bowtie structures with shorter gap distances might provide even larger detection sensitivities. The case reported here could thus be considered as the lowest detection limit.

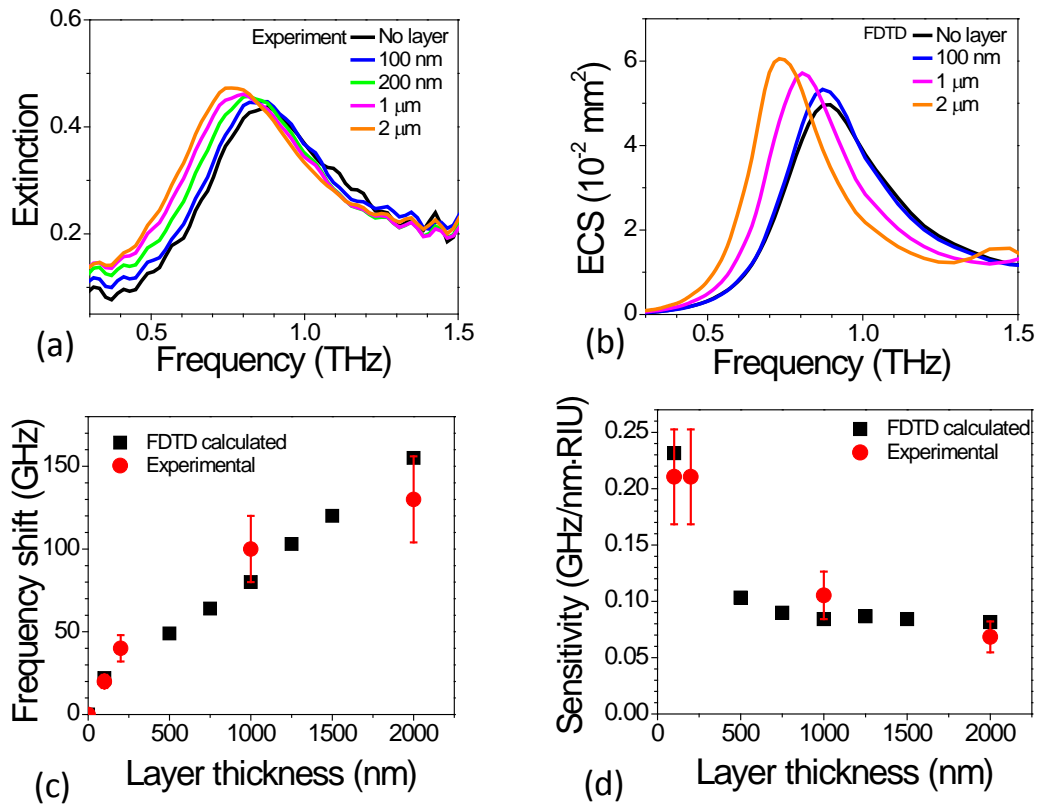

Fig. 2. a) Extinction, defined as (1-Transmission), of a collection of doped silicon bowtie antennas as a function of the thickness of the deposited $SiO_2$ layer; b) FDTD calculated extinction cross section (ECS) of an individual doped silicon bowtie antenna covered with $SiO_2$ layers of different thickness ; c) Shift of the resonance frequency for the same situation as in a) and b) as a function of the top layer thickness, as obtained from experiments (red circles) and from FDTD calculations (black squares); d) Sensitivity of the bowtie antenna expressed in frequency shift per nanometer of film thickness, estimated from the experimental (red circles) and calculated (black squares) shift of the resonance.

In order to quantify the sensitivity of the bowtie antennas to the $SiO_2$ coverage, we have measured the frequency shift of the resonance as a function of the layer thickness (red circles) and compared it to the simulations (black squares), as shown in Fig. 2(c). The error bars on the experimental points cover the measurement uncertainties as well as the sample to sample fluctuations. We observe that the experimental resonant frequency shift does not increase linearly and tends to saturate for thicker layers. Even though the calculated resonance frequency shifts do not exhibit the saturation for exactly the same thickness as in the experiments, the qualitative behavior is reproduced fairly well. This behavior is due to the fact that the electromagnetic field is confined to the proximity of the plasmonic structure and decays over micron length scales from the interface. The sensitivity $S$ of the technique towards the properties of the layer can be expressed as $S = \Delta v/(\Delta n \cdot t)$ where $\Delta v$ is the frequency shift of the resonance, $\Delta n$ is the refractive index difference between air and the covering layer and $t$ the thickness of the covering layer. The sensitivity of the localized surface plasmon resonance of the bowtie antenna is plotted *versus* the layer thickness in Fig. 2(d). $S$ is largest for the thinnest layers, owing to the field confinement near the surface of the structure. This behavior gives the indication that an eventual THz sensor based on silicon plasmonic antennas would perform particularly well in the range of thickness between 100 nm and 2000 nm. The detection of larger layer thickness is possible, even though the sensitivity to variations in the thickness is lower in this case. We note the ability of THz plasmonic antennas to detect layers with low optical thickness $n \cdot t$. This can be particularly interesting for the detection of organic and biological layers, which are usually characterized by low refractive indices.

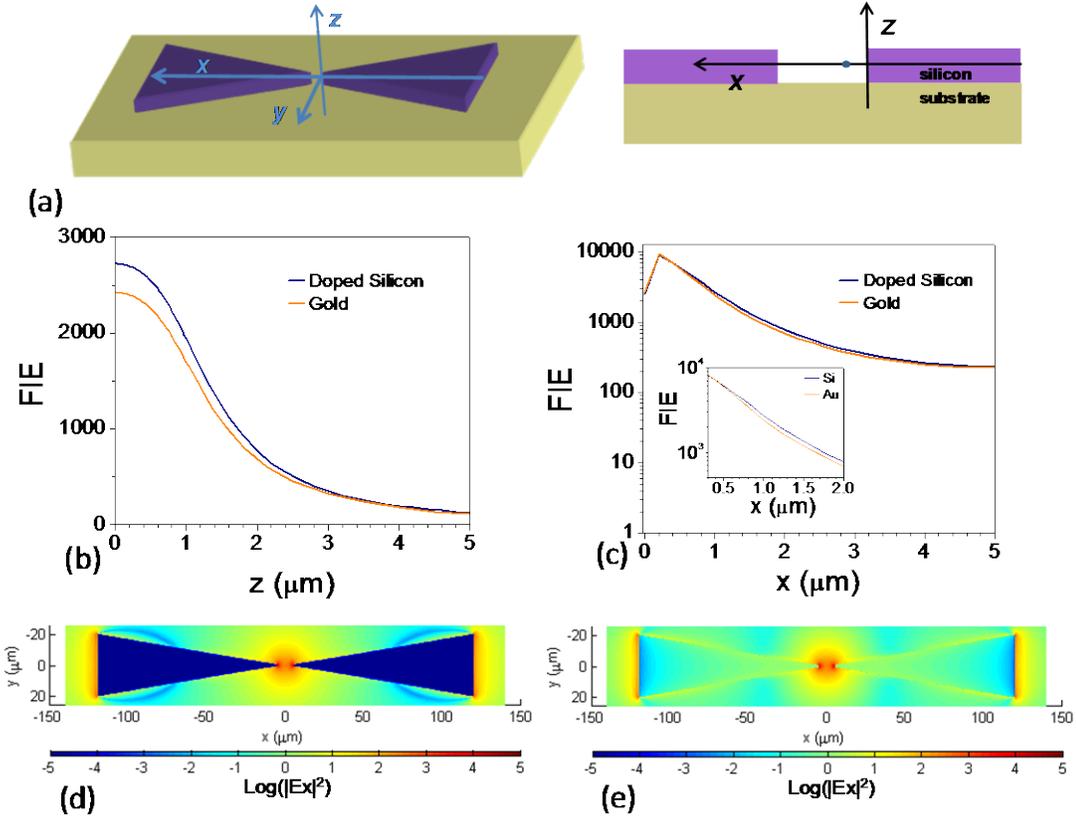

Fig. 3. a) Schematics of the bowtie antenna in top view (left) and cross sectional view (right). The direction of the relevant axes are indicated. The blue dot on the cross sectional view illustrates the position y=z=0 and x=1 μm; b) Field intensity enhancement (FIE) obtained with FDTD calculations for a bowtie antenna without coating along the z direction at x=1 μm. The case of a silicon antenna (blue line) is compared to that of a gold antenna with the same geometry (red line). The position z=0 μm corresponds to the surface of the silicon antenna; c) FIE obtained from FDTD calculations in a bowtie antenna without coating along the x direction, at z=0 μm. The position z=0 μm is located at the middle of the antenna height, *i.e*, at 0.75 μm from the interface with the quartz substrate. The position x=0 μm corresponds to the tip of one of the monomers of the bowtie antenna. Doped silicon antennas (blue line) and gold antennas (red line) with the same geometry are compared. The inset is a closed view of the decay of the field for x between 300 nm and 2 μm; d) Field profile for an antenna made of gold calculated by FDTD in the middle plane of the antenna at 0.8THz, the colormap is plotted in logarithmic scale; e) Field profile for an antenna made of doped silicon calculated by FDTD in the middle plane of the antenna at 0.8THz, the colormap is plotted in logarithmic scale.

## 4. Near-fields

The sensitivity of the bowtie structures to thin layers is related to the local electromagnetic field induced around the THz antenna, so it is worthwhile to analyze the spatial distribution of the local field. Figure 3 shows the calculated field profiles in a cross section of the antennas in the *x-z* plane, as schematically indicated in Fig. 3(a). We define the *x* axis to be located in the middle of the height of the bowtie antenna, *i.e.*, at 0.75 μm above the interface silicon/substrate, pointing towards the gap of the antenna. We define the field intensity enhancement (FIE) by the ratio of the field intensity to the incident field intensity. Figure 3(b) represents the decay of the FIE in the vertical direction, normal to the substrate. The FIE calculations shown in Fig. 3(b) were performed at y=0 and *x*=1 μm. This position has been chosen in the gap of the antenna, close to the tip. It has been checked that calculations at other locations closer to the tip yield similar results: the FIE decays over a few micrometers. We compare the FIE in the proximity of the bowtie antennas made of doped silicon to that produced in a metallic (gold) antenna with the same dimensions. According to the results shown in Fig. 3(b), bowtie antennas made of doped silicon show a slightly larger field enhancement in the transverse direction than those made of gold for micrometer distances from the antenna surface. Therefore doped semiconductors should in principle show better potential for THz sensing of nanometric layers. The confinement of the field enhancement is similar in both materials as derived from the field distribution shown in Fig. 3 (b) and (c). Figure 3(c) displays the FIE in the middle plane of the antenna, along the *x* axis from the tip of the antenna inwards the gap. We note that the FIE is more than $10^2$ over the whole bowtie gap, and increases close to the tips. The maximum of FIE appears for *x*=200 nm. This is explained by the fact that the mesh size in this direction is 200 nm. The maximum field enhancement is expected to occur at the triangle tip and the FIE decreases further away from the tip. The pronounced FIE at the apex of the bowtie monomers is due to the sharpening of the structure, which produces a lightning rod effect [26]. Stronger field enhancements leading to larger sensitivity can be expected for smaller antenna gaps when the two segments of the bowtie antenna are more strongly coupled [22].

5. **Conclusion**

We have shown that very thin inorganic films (3750 times thinner than the free space wavelength) can be detected with THz radiation when the films are deposited on top of semiconductor bowtie antennas. Thin films of $SiO_2$ and $TiO_2$ have been deposited with thicknesses varying from 100 nm to 2 µm. The detection is based on a red-shift of the localized plasmon resonance induced by the presence of the layer with respect to the pristine bowtie antenna. This red-shift is more pronounced when the optical thickness of the layer deposited increases, and for a given value of the refractive index, it saturates with increasing film thickness. The sensitivity to the thin film is directly related to the local electromagnetic field distribution in the proximity of the antenna. Good agreement between the experiments and the FDTD simulations is obtained for the resonance frequencies in all cases. As organic or biological constituents have similar optical properties to the films studied in this work, the demonstrated ability to detect very thin, low refractive index layers paves the way for improved biological sensing applications at THz frequencies. A promising application can be the interaction with cells and microorganisms owing to the size of THz hot spots. For other applications, a detection scheme for enhanced spectroscopy such as presented in this article could be applied for gas identification if a surface with plasmonic antennas is inserted in a flow cell with the gas to be identified. Gases with specific absorption lines will couple to the plasmonic resonance. For explosive detection, the deposit of small amounts of powder on the plasmonic antennas could allow detection of smaller amounts than conventional THz spectroscopy. This work demonstrates that it is possible to significantly reduce the sensing volume. The distribution of field enhancement around the bowtie antenna indicates that this sensing volume can be reduced even further when the material to detect is placed in the regions of large field enhancement (mainly the gap of the antenna), or if a reduced amount of antennas is used down to the limit of single antenna spectroscopy.

**Acknowledgements:**

This work was supported by the European Community's 7[th] Framework Programme under grant agreement n[o] FP7-224189 (ULTRA project, http://www2.teknik.uu.se/Ultratc) and is part of the research program of the "Stichting voor Fundamenteel Onderzoek der Materie (FOM)", which is financially supported by the "Nederlandse Organisatie voor Wetenschappelijk Onderzoek (NWO)". This work was also supported by the European FP7 project "Nanoantenna" (FP7-HEALTH-F5-2009-241818-NANOANTENNA).